\documentclass[twocolumn,aps,prc,tightenlines,floats,floatfix,eqsecnum,preprintnumbers,nofootinbib]{revtex4}

\def\sp{\kern +3pt}
\def\sm{\kern -3pt}
\def\spQ{\kern +6pt}

\def\bea{\begin{eqnarray}}
\def\eea{\end{eqnarray}}

\def\sfrac#1#2{{\textstyle \frac{#1}{#2}}}

\newcommand{\ket}[1]{|#1\rangle}

\def\be{\begin{equation}}
\def\ee{\end{equation}}
\def\ba{\begin{eqnarray}}
\def\ea{\end{eqnarray}}

\usepackage{graphics}
\usepackage{graphicx}
\usepackage{epsf}
\usepackage{amsmath}
\usepackage{amssymb}

\begin{document}

\phantom{0}
\vspace{-0.2in}
\hspace{5.5in}

\preprint{}

\vspace{-1in}

\title
{\bf What is the role of the meson cloud
in the
$\Sigma^{*0} \to \gamma \Lambda$ and
$\Sigma^\ast \to \gamma \Sigma$ decays?}
\author{G.~Ramalho$^1$ and
K.~Tsushima$^1$}
\vspace{-0.1in}

\affiliation{
$^1$International Institute of Physics, Federal
University of Rio Grande do Norte, Avenida Odilon Gomes de Lima 1722,
Capim Macio, Natal-RN 59078-400, Brazil
}

\vspace{0.2in}
\date{\today}

\phantom{0}

\begin{abstract}
We study the effect of the meson cloud dressing
in the octet baryon to decuplet baryon electromagnetic transitions.
Combining the valence quark contributions from
the covariant spectator quark model
with those of the meson cloud estimated based on the flavor SU(3)
cloudy bag model, we calculate the transition magnetic form factors
at $Q^2=0$ ($Q^2=-q^2$ and $q$ the four-momentum transfer),
and also the decuplet baryon electromagnetic decay widths.
The result for the
$\gamma^\ast \Lambda \to \Sigma^{\ast 0}$ decay width
is in complete agreement with the data,
while that for the $\gamma^\ast \Sigma^+ \to \Sigma^{\ast +}$
is underestimated by 1.4 standard deviations.
This achievement 
may be regarded as a significant advance
in the present theoretical situation.
\end{abstract}

\vspace*{0.9in}  
\maketitle

\section{Introduction}

One of the most interesting challenges
in hadronic physics is to study
the internal structure of baryons and mesons.
A microscopic understanding
of the transition between the hadronic states
is also very important.
Although it is generally accepted that the internal structure
of hadrons and the dynamics of quarks and gluons,
are described by quantum chromodynamics (QCD),
one has to rely on some effective degrees of freedom
in the nonperturbative low $Q^2$ region
such as constituent quarks which form baryon cores
with meson cloud excitations~\cite{Burkert04a,NSTAR}.
Although there exist some works
which attempted to treat 
the meson cloud explicitly as the $q \bar q$ excitations
in the so-called
{\it unquenched quark models}~\cite{Geiger97,Capstick00,Bijker09,Capstick08},
most of the phenomenological models treat the meson cloud
using pointlike meson excitations.

Particular examples of very interesting studies
may be the electromagnetic transitions between
an octet baryon $B$ (spin 1/2) and
a decuplet baryon $B'$ (spin 3/2),
$\gamma^\ast B \to B'$,
and the $B'$ electromagnetic decay reactions, $B' \to \gamma B$.
There are theoretical predictions
for the $\gamma^\ast B \to B'$ transition
magnetic moments based on 
quark models~\cite{Lipkin73,Koniuk80,Darewych83,Warns91,Sahoo95,Wagner98,Bijker00},
including quark models with
meson cloud dressing~\cite{Kaxiras85,Lu97,Sharma10,Yu06},
Skyrme and soliton models~\cite{Schat96,Haberichter97},
large $N_c$ limit~\cite{Lebed11},
QCD sum rules~\cite{Wang09,Aliev06}, and
chiral perturbation theory~\cite{Butler93a}.
There are also some results
from lattice QCD~\cite{Leinweber93,Arndt04}.
One of the strong motivations to study the $\gamma^\ast B \to B'$ reactions
is to clarify the role of the meson cloud dressing,
which is of fundamental importance, as was demonstrated by
the $\gamma^\ast N \to \Delta$
reaction~\cite{Burkert04a,Pascalutsa07,NDelta,NDeltaD,Lattice,LatticeD}.
The data, except for the $\gamma^\ast N \to \Delta$  reaction,
namely the $\Sigma^{\ast 0} \to \gamma \Lambda$
and $\Sigma^{\ast +} \to \gamma \Sigma^+$ decay widths,
have become available only recently~\cite{PDG,Taylor05,Keller11a,Keller11b}.
In general, most of the model predictions
significantly underestimate the  
data, particularly
those for the $\Sigma^{\ast +} \to \gamma \Sigma^+$ decay width
(see Ref.~\cite{Oct2Dec} for a more detailed discussion).

In our previous work~\cite{Oct2Dec} we studied the
$\gamma^\ast B \to B'$ reactions using
a covariant constituent quark model,
complemented by the pion cloud effects
extrapolated by the $\gamma^\ast N \to \Delta$ reaction
based on an SU(3) symmetry.
The pion cloud effects were included
in the leading order, namely, they included only the processes with
the direct photon coupling to the pion.
The electromagnetic transition form factors calculated
were decomposed into the valence quark and pion cloud contributions.
We concluded that the pion cloud effects
could help to explain satisfactorily the
$\gamma^\ast N \to \Delta$ data,
but only partially help to explain the data for the
$\gamma^\ast \Lambda \to \Sigma^{\ast 0}$
and  $\gamma^\ast \Sigma^+ \to \Sigma^{\ast +}$ reactions.
Therefore, the other effects, such as
the contributions from the heavier mesons
like the kaon, and alternative higher order processes
involving the meson cloud, may be relevant
to explain the experimental decay widths.
The next-order processes to be included in this study
are the processes which one photon couples
to the intermediate baryon states
while one meson is in the air.
In addition, the heavier mesons
to be taken into account are the kaon and eta meson,
the next lighter mesons to the pion.

In a model with pointlike quarks 
the photon coupling with the intermediate 
baryon states is not expected to be important
for the meson cloud contributions,
since the octet to the decuplet electromagnetic transitions
are dominated by the magnetic interactions,
and the quark anomalous magnetic moments vanish
for the pointlike quarks.
However, in a constituent quark model
like the one we use in this study,
the covariant spectator quark model,
the octet-decuplet electromagnetic transitions
are dominated by the mechanisms with a quark spin flip
(magnetic type interactions).
Therefore, the valence quark contributions
may be very important when
the quark anomalous magnetic moments are significant.
Furthermore, we can expect important meson cloud contributions
from the intermediate octet-decuplet baryon electromagnetic transitions  
while one meson is in the air  (photon-vertex correction),
since these mechanisms also depend on the
quark anomalous magnetic moments.

\begin{figure}[t]
\vspace{.4cm}
\includegraphics[width=2.6in]{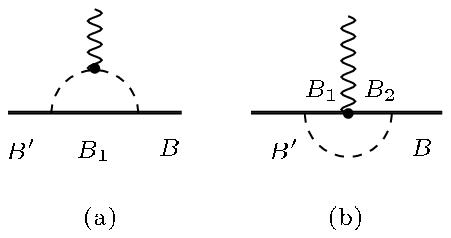}
\caption{\footnotesize
Meson cloud contributions
for the electromagnetic transition form factors.
Between the initial octet ($B$) and
final decuplet ($B'$) baryon states,
there are several possible intermediate baryon states:
(a) $B_1$; 
(b)
$B_1$ and $B_2$.
Depending on the meson $M$, 
meson-baryon states ($M$-$B_1$ and $M$-$B_2$)
may arise, 
where $B_1$ and $B_2$ are the   
octet and decuplet baryon states in this study. 
}
\label{figMesonCloud}
\end{figure}

In this work we improve the
calculation of the meson cloud contributions
for the $\gamma^\ast B \to B'$ transitions
made in the previous work~\cite{Oct2Dec},
and predict the corresponding
electromagnetic decay widths (determined at $Q^2=0$).
The improvements are the following:
i) inclusion of the photon coupling
to the intermediate baryon states;
ii) inclusion of the effects of
the heavier meson clouds, kaon and eta meson,
besides the pion.
As in the previous work, the transition form factors
can be decomposed into the valence quark and meson cloud
contributions. 
The processes included as the meson cloud contributions
in this work are depicted in Fig.~\ref{figMesonCloud},
in terms of the meson and baryon degrees of freedom.

We will conclude that the effects of the
intermediate baryon states combined with
the kaon cloud, improve the
agreement of our model 
with the experimental data.

The valence quark contributions
are estimated based on the covariant spectator quark
model~\cite{ExclusiveR,Nucleon,Nucleon2,OctetFF,Omega}
as in the previous work.
Thus, the baryons are described as
three-quark systems.
The valence quark contributions for the
transition form factors are calculated
using the octet and decuplet baryon wave functions
and the quark electromagnetic current
of the model,
determined in the previous works.

To describe the meson cloud contributions
for the octet-decuplet baryon electromagnetic transitions
we need a microscopic model 
to describe the virtual meson-baryon states.

Contrary to the valence quark contributions
that dominate in the large $Q^2$ region,
the meson cloud effects are
long-range processes,
and are known to be of crucial to explain
the transition helicity amplitudes
and form factors in the low $Q^2$
region~\cite{Burkert04a,NSTAR}.
To incorporate the meson cloud effects,
we use the cloudy bag model
(CBM)~\cite{Lu97,Thomas84,Thomas83,TsushimaCBM,Umino93}
which treats the mesons
as pointlike particles to describe
the meson cloud dressing
in the static approximation for the baryons.
All such approximations have been
practiced well in the past within the CBM,
and may be regarded as under control.

Although the CBM framework differs from
the covariant spectator quark model
for the treatment of the valence quarks,
the CBM can be used as an effective description
of the long-range physics of meson cloud dressing.
The possible conflict between the two models
e.g., the lack of the explicit covariance 
and the limitation of the applicability
for the large $Q^2$ region in the CBM,
can be overcome in a proper manner,
since one can define a covariant extension
of the model based on covariant parametrizations
for the meson cloud contributions that
are equivalent with the CBM result at $Q^2=0$.
See for instance Refs.~\cite{Delta1600,LambdaSigma0,OctetMedium},
where meson cloud contributions were estimated in different reactions. 
In addition, the merit of using the CBM is
that the model is based on 
SU(3) (SU(6) flavor-spin) symmetry 
and chiral symmetry.

The explicit calculation of the meson cloud
contributions considered in this study
requires two kind of mechanisms.
The first mechanism is the photon coupling with the meson,
and for this, we use a formalism similar to that
applied in Ref.~\cite{Oct2Dec}.
However, in the present study we take into account
the explicit dependence of the baryon
and meson (pion, kaon and eta) masses.
[Previously, we used SU(3) symmetry
for the baryon masses, and only the pion cloud was included.]
The second mechanism is the photon coupling with the intermediate baryon
states, which is more delicate and model dependent,
since this requires an estimate of all the
intermediate octet-octet, octet-decuplet
decuplet-octet and decuplet-decuplet transition form factors at $Q^2=0$.
The corresponding expressions are derived
in the CBM framework, 
but since we describe the
valence quark cores with the covariant
spectator quark model, it is necessary to reinterpret
the CBM quark magnetic moments in terms of
those calculated by the spectator quark model.
This will be done using SU(3) symmetry
to be explained in detail later.

Finally, the results from the CBM are normalized
by the pion cloud contribution
obtained in the covariant spectator quark
model for the $\gamma^\ast N \to \Delta$ transition~\cite{NDelta},
under the assumption that the pion cloud is the dominant
meson cloud contribution.
With this procedure also used  
in the previous work~\cite{Oct2Dec},
we preserve the parametrization of the
covariant spectator quark model for the core,
and estimate the effects of the meson cloud
for the other octet to decuplet transitions,
as well as the kaon and eta clouds
for the $\gamma^\ast N \to \Delta$ reaction.

This article is organized as follows:
in Sec.~\ref{secFormalism} we explain
the decomposition of the valence and meson cloud
contributions for the transition form factors.
In Sec.~\ref{secValence} we review
the formalism associated with the
valence quark contributions for the form factors,
and express the results in terms of
{\it effective} quark magnetic moments,
which are also necessary for the
calculation of the meson cloud effects.
In Sec.~\ref{secMeson} we present the formalism
associated with the meson cloud dressing.
The results are presented in Sec.~\ref{secResults},
while the final conclusions are given in Sec.~\ref{secConclusions}.

\section{Formalism}
\label{secFormalism}

Next, we discuss briefly the formalism necessary 
to describe the valence quark contributions,
as well as the mechanism of the meson cloud dressing.

In the covariant spectator quark model, 
baryons are treated as
three-quark systems~\cite{ExclusiveR,Nucleon,Nucleon2,OctetFF,Omega}.
The electromagnetic interactions with the baryons
are described by the photon coupling with
the constituent quarks in the relativistic impulse approximation,
and the quark electromagnetic structure is
represented 
in terms of the quark form
factors parameterized by a vector meson dominance
mechanism~\cite{Nucleon,Omega}.
The parametrization of the quark current,
calibrated previously 
in the studies of the nucleon form factors~\cite{Nucleon}
and by the lattice QCD data for the decuplet baryons~\cite{Omega},
encodes effectively the gluon
and quark-antiquark 
substructure of the constituent quarks.
The baryon wave functions derived
from the SU(6)$\otimes$O(3) structure,
are written in terms of an off-shell quark, that is free
to interact with the photon fields,
and two on-shell quarks.
Integrating over the quark-pair degrees
of freedom, we reduce the three-quark baryon state
to a quark-diquark state,
where the diquark can be represented as
an on-shell spectator particle with an effective
mass of $m_D$~\cite{Nucleon,Nucleon2,Omega}.

Under the assumption that each baryon system can be described
by the wave function with an S-state configuration
for the quark-diquark system in the first approximation,
we calculated the valence quark contributions for
the magnetic form factors $G_M^B$ in the previous work~\cite{Oct2Dec},
where the upper index $B$ 
labels the contributions from the quark core (bare),
using the wave functions from Refs.~\cite{OctetMedium,Omega}.
(See Ref.~\cite{Oct2Dec} for more details.)
Contributions from the electric and Coulomb quadrupole form factors
appear only beyond the S-state approximation for the
decuplet baryon wave functions.
However, their contributions are expected to be small
(small orbital angular momentum admixtures)~\cite{NDeltaD,LatticeD},
and thus they are neglected in this work.

As mentioned already, the constituent quarks considered 
in this work, have internal structure,
and the  structure is encoded in a vector meson dominance
parametrization~\cite{Nucleon,OctetFF,Omega},
that includes effectively, among other effects,
the meson cloud dressing of the quarks.
However, it should be emphasized that
there are meson cloud effects that cannot be included
in the constituent quark structure, 
such as the process of meson exchange between 
the different quarks inside the baryon,
which cannot be reduced
to a simple diagram of a quark dressing.
Processes of this kind, have to be represented
at the hadronic level
(meson and baryon states) as the diagrams
shown in Fig.~\ref{figMesonCloud}.
Thus, the meson cloud in this study, is regarded
as a process of one meson exchange between
the different quarks inside
the baryon~\cite{OctetFF}.
Since the meson cloud dressing can
appear in two independent mechanisms
(self dressing of the quarks and the others)
there is  no double counting.
In summary, besides the contributions from
the valence quark core calculated using the
quark electromagnetic form factors,
there are meson cloud effects that
have to be taken into account
in the electromagnetic transitions between
the baryon states.
Those meson cloud effects are the
main focus of the present work.

From the discussions made previously,
we conclude that 
the magnetic transition
form factors ($G_M$) can be represented as
the sum of the valence quark ($G_M^B$)
and meson cloud ($G_M^{\rm MC}$)
contributions in the present approach:
$G_M= G_M^B + G_M^{\rm MC}$.
In particular, for the study of the
baryon decuplet decay widths,
we need to consider only the case $Q^2=0$.
Thus, we can write
\ba
G_M(0)= G_M^B(0) + G_M^{\rm MC}(0).
\ea

As mentioned already, the meson cloud contribution
for the octet to decuplet transition,
can be decomposed in the two processes displayed
in Fig.~\ref{figMesonCloud},
for the first and second order,
classified by the number of the baryon propagators.
The diagram (a) represents the direct coupling
of a photon with the intermediate state meson
(first order, one baryon propagator).
The diagram (b) represents the direct coupling
of a photon with the intermediate state baryons
(second order, two baryon propagators).
One can then decompose $G_M^{\rm MC}(0)$ into the contributions
from the diagrams (a) and (b):
\ba
G_M^{\rm MC}(0)=G_M^{\rm MCa}(0) +  G_M^{\rm MCb}(0).
\ea
The meson cloud contributions corresponding to each diagram
(a) and (b), can be further decomposed
into the pion, kaon, and eta cloud contributions.

As in the previous work~\cite{Oct2Dec}, in order to
keep the parametrization of the covariant
spectator quark model,
we regularize the results for the pion cloud contribution
by that from the
covariant spectator quark model
for the $\gamma^\ast N \to \Delta$ reaction,
\ba
G_M^{{\rm MC} \pi}(0) = 3 \lambda_\pi,
\ea
where $\lambda_\pi =0.441$ defines the strength
of the pion cloud effect~\cite{NDeltaD}.
In this procedure we assume that the pion cloud
is the dominant meson cloud effect
in the $\gamma^\ast N \to \Delta$ reaction.
Later we will see that this assumption is indeed justified.
For the other octet to decuplet transition reactions,
and also for all the meson clouds considered
in the present work, we use the relation,
\ba
G_M^{\rm MC} (0) = f_{BB'} (3 \lambda_\pi),
\label{eqFBBp}
\ea
where the factor $f_{BB'}$ contains
the pion, kaon and eta meson cloud contributions
from the both diagrams (a) and (b) in Fig.~\ref{figMesonCloud}.
The calculation of the coefficient $f_{BB'}$
will be explained in Sec.~\ref{secResults}.

The diagram (b) includes in the intermediate states,
the octet-octet, octet-decuplet, decuplet-octet
and decuplet-decuplet baryon electromagnetic transitions.
Therefore, to estimate the contributions
from the possible intermediate baryon state transitions,
we need to calculate all the
corresponding transition magnetic form factors
at $Q^ 2=0$.
One can in principle calculate them in
the covariant spectator quark models for this purpose,
however, the explicit estimates corresponding to all
the intermediate baryon state transitions
would be complex and tedious. 
Therefore, for the diagram (b) we use
the estimate made in the CBM/SU(6) framework,
where all the intermediate
state contributions can be related with the valence quark magnetic moments.
In order to relate the CBM/SU(6) quark magnetic moments
with the anomalous magnetic moments
in the covariant spectator quark model,
we will start by reviewing
the expressions used for the valence quark
contributions for the octet to decuplet
electromagnetic transition form factors.

\section{Valence quark contributions}
\label{secValence}

In the covariant spectator quark model
the valence quark contributions for
the form factors are calculated
using the octet and decuplet baryon wave functions,
and the constituent quark current.
The quark current has the general form~\cite{Nucleon,Omega},
\ba
j_q^\mu(Q^2) = j_1(Q^2) \gamma^\mu
+ j_2(Q^2) \frac{i \sigma^{\mu \nu} q_\nu}{2M_N},
\ea
where $j_i$ $(i=1,2)$ are the quark form factors that
can be parametrized in terms of a vector dominance
mechanism.
The form factors $j_i$ can also be decomposed
in the quark-isoscalar, quark-isovector and
strange-quark components.
The details can be found in Refs.~\cite{Oct2Dec,Nucleon,OctetFF,Omega},
but are not important for the present discussion,
since we are considering the $Q^2=0$ case,
where $j_1(0)= e_q$ and $j_2(0)= e_q \kappa_q$.
The last equation defines the quark anomalous moment ($\kappa_q$)
in the covariant spectator quark model formalism.

To calculate the transition form factors,
we project the operator $j_i$ on the
mixed anti-symmetric ($\ket{M_A}$)
and mixed symmetric ($\ket{M_S}$)
components of the octet and
(fully symmetric) decuplet $\ket{B'}$ flavor states:
\ba
& &
j_i^A= 3 \left< B' | j_i | M_A \right>, \\
& &
j_i^S= 3 \left< B' | j_i | M_S \right>.
\ea
More details can be found in Refs.~\cite{Oct2Dec,OctetFF,Omega}.
Note that, for the octet to decuplet baryon transitions
only the components $j_i^S$ (isovector) are relevant.
Finally, the magnetic form factor can be written~\cite{Oct2Dec} as,
\ba
G_M^B = \frac{2 \sqrt{2}}{3}
\sqrt{\frac{2}{3}} \bar f_v {\cal I},
\label{eqGM0}
\ea
with
\ba
\bar f_v =
\frac{2 M_B}{M_{B'} + M_B}
\left\{
\frac{j_1^S}{\sqrt{2}}  +
\frac{M_{B'} + M_B}{2 M_N} \frac{j_2^S}{\sqrt{2}}
\right\},
\label{eqFv}
\ea
where the coefficients
$\frac{1}{\sqrt{2}} j_i^S$
can be found in Ref.~\cite{Oct2Dec},
and ${\cal I}$ is the overlap integral
between the octet ($\psi_B$) and decuplet ($\psi_{B'}$)
radial wave functions (see details also in Ref.~\cite{Oct2Dec}).
The radial wave functions $\psi_B$ and $\psi_{B'}$
are scalar functions of the baryon and
diquark momenta~\cite{Oct2Dec,Omega,OctetFF,OctetMedium}.
In Eqs.~(\ref{eqGM0}) and~(\ref{eqFv}) $G_M^B$, $\bar f_v$ and
${\cal I}$ are exclusive functions of $Q^2$.

Now, we focus again on the $Q^2=0$ case.
In this case we can write the factor $\bar f_v$ in a more
compact form, defining the
{\it effective} quark magnetic moment 
of the transition
$\gamma^\ast B \to B'$ as
\ba
\hat \mu_q =
\frac{2 M_{B}}{M_{B'} + M_B} +
\frac{M_B}{M_N} \kappa_q. 
\label{eqMU}
\ea
Note that the expression for $\hat \mu_q$
is reduced  to the usual form,
$\mu_q = (1 + \kappa_q)$,
in the limit $M_{B'}=M_B= M_N$.
However, $\hat \mu_q$ now depends on the masses
of the ``submultiplets''
($M_{B'}$ and $M_B$ in the $\gamma^* B \to B'$ transition).
We keep this dependence in mind, but
suppress the indices $B$ and $B'$ in $\hat \mu_q$
for simplicity.

The explicit expressions for $\bar f_v$ in terms of $\hat \mu_q$
are presented in Table~\ref{tableJS}.
In particular, we can express the reactions involving
the $\Sigma$ and $\Xi$ as,
\ba
\bar f_v =
\frac{1}{6} (2 \hat \mu_u - \hat \mu_d + 2 \hat \mu_s) 
+
\frac{1}{6} (2 \hat \mu_u + \hat \mu_d) t_3,
\label{eqFV}
\ea
where $t_3= J_3$ for $\Sigma$
and $t_3 = \tau_3$ for $\Xi$.
The matrices $J_3={\rm diag}(1,0,-1)$ and
$\tau_3= {\rm diag}(1,-1)$,
are, respectively, isospin-1 and isospin-1/2
operators that act on the isospin states
of the baryons $B$ and $B'$.

\begin{table*}[t]
\begin{tabular}{l   cc cc cc}
\hline
\hline
    && $\bar f_v$ &&
$G_M^B$ &&$G_M^B$(CBM)\\
\hline
\hline
$\gamma^* p \to \Delta^+$  &&
$\frac{1}{3} ( 2 \hat \mu_u + \hat \mu_d)$ &&
$A \frac{1}{3} (2 \mu_u + \mu_d)$ &&
$A \bar \mu_u$
  \\
$\gamma^* n \to \Delta^0$  &&
$\frac{1}{3} (2 \hat \mu_u +  \hat \mu_d)$ &&
$A \frac{1}{3} (2 \mu_u + \mu_d)$ &&
$A \bar \mu_u$
\\[.3cm]
$\gamma^* \Lambda \to \Sigma^{\ast 0}$ &&
$\sqrt{\frac{3}{4}} \frac{1}{3} (2 \hat \mu_u +  \hat \mu_d)$ &&
$ \sqrt{\frac{3}{4}} A
\frac{1}{3} (2 \mu_u +  \mu_d)$ &&
$ \sqrt{\frac{3}{4}} A
 \bar \mu_u $
\\ [.3cm]
$\gamma^* \Sigma^+ \to \Sigma^{\ast +}$  &&
$\frac{1}{3} (2 \hat \mu_u +\hat \mu_s)$ &&
$A
\frac{1}{3}
(2 \mu_u + \mu_s )$ &&
$A
\frac{1}{3}
(2 \bar \mu_u + \mu_s )$
\\
$\gamma^* \Sigma^0 \to \Sigma^{\ast 0}$  &&
$\frac{1}{6} ( 2 \hat \mu_u -\hat \mu_d + 2\hat \mu_s)$ &&
$A
\frac{1}{6}
( 2\mu_u - \mu_d + 2 \mu_s )$  &&
$A
\frac{1}{3}
( \bar \mu_u + 2 \mu_s )$
\\
$\gamma^* \Sigma^- \to \Sigma^{\ast -}$ &&
$\frac{1}{3} ( - \hat \mu_d +  \hat \mu_s)$ &&
$A \frac{1}{3} ( - \mu_d +  \mu_s)$ &&
$A
\frac{1}{3}
( -\bar \mu_u +  \mu_s )$
  \\[.3cm]
$\gamma^* \Xi^0 \to \Xi^{\ast 0}$ &&
$\frac{1}{3} (2\hat \mu_u + \hat \mu_s)$ &&
$A \frac{1}{3} (2 \mu_u + \mu_s)$ &&
$A
\frac{1}{3}
(2 \bar \mu_u +  \mu_s )$
\\
$\gamma^* \Xi^- \to \Xi^{\ast -}$ &&
$\frac{1}{3} (- \hat \mu_d +  \hat \mu_s)$ &&
$A \frac{1}{3} (- \mu_d +  \mu_s)$ &&
$A
\frac{1}{3}
( -\bar \mu_u + \mu_s )$
\\[0.02in]
\hline
\hline
\end{tabular}
\caption{
Coefficients $j_i^S$ ($i=1,2$), $\bar f_v$
and valence quark contributions for the $G_M$ form factors.
In the expressions for $G_M^B$, one has
$A= \frac{2 \sqrt{2}}{3}$.
}
\label{tableJS}
\end{table*}

The contributions from the valence quarks
for the form factors
can also be estimated by the SU(6) quark
model in terms of the quark magnetic moments $\mu_q$.
The results are expressed in terms
of the $u, d$ and $s$  quark magnetic moments,
$\mu_u,\mu_d$ and $\mu_s$, respectively.
The CBM uses the spin-flavor SU(6) wave functions,
but calculates the values of $\mu_q$
using the CBM (MIT bag) formalism.
Since usually $\mu_u  \equiv \mu_d$ in the CBM
(reflecting the quark masses used, $m_u=m_d$),
we use $\bar \mu_u$ to represent either $\mu_u$ or $\mu_d$.

Note that the definitions of the
quark magnetic moments discussed here do not
include the quark charges,
contrarily to the convention used
for instance in naive quark models~\cite{PDG2,Cloet02}.

The results for $G_M^B$ (quark core contributions)
from the CBM are presented
in the last column in Table~\ref{tableJS}.
For the $\gamma^\ast N \to \Delta$ reaction
the result is~\cite{Pascalutsa07,Koniuk80}
\ba
G_M^B(0)= \frac{2 \sqrt{2}}{3} \bar \mu_u,
\hspace{1cm} \mbox{(SU(6))},
\label{eqSU6}
\ea
where $\bar \mu_u = \mu_p$, the proton magnetic moment
in the SU(2) limit\footnote{For simplicity
we ignore the factor $\sqrt{\frac{M_N}{M_\Delta}}$
that transforms magnetic moment $\mu_{N\Delta}$
into the corresponding form factor $G_M(0)$.
This simplification has no consequence
in the present work, since to
identify the results of the CBM
and those of the spectator quark model,
global factors are not important.}.
In this limit also 
$\mu_n = -\frac{2}{3} \bar \mu_u $.

In order to compare the results of the
covariant spectator quark model
with those of the CBM, we need to
relate the spectator model quark magnetic moments 
with those of the CBM/SU(6). 
Motivated by Eq.~(\ref{eqSU6}),
and taking into account that the structure of $G_M^B$
in the covariant spectator quark model, 
given by Eqs.~(\ref{eqGM0}) and (\ref{eqFv}),
we define $\mu_q \equiv \sqrt{\frac{2}{3}} \hat \mu_q {\cal I}(0)$, or
\ba
\mu_q = \sqrt{\frac{2}{3}}
\left\{
\frac{2 M_{B}}{M_{B'} + M_B} +
\frac{M_B}{M_N} \kappa_q \right\}
{\cal I}(0),
\label{eqMUb}
\ea
for the covariant spectator quark model.

With the above identification of the
quark magnetic moments, we can write the
$\gamma^\ast N \to \Delta$ magnetic form factor
in the covariant spectator quark model as
\ba
G_M^B(0)= \frac{2 \sqrt{2}}{3}
\frac{1}{3}(2 \mu_u + \mu_d),
\hspace{1.cm} \mbox{(Spectator)}.
\label{eqGBsp}
\ea
Note the similarity between
Eqs.~(\ref{eqSU6}) and~(\ref{eqGBsp}).
If we replace
$\sfrac{1}{3}(2\mu_u + \mu_d) \to \bar \mu_u$
as in the SU(2) symmetric case, the two equations are equivalent.
The expressions for the other octet to decuplet
transitions are presented in Table~\ref{tableJS},
with $A= \frac{2\sqrt{2}}{3}$.
Also for the other reactions 
the expressions from the covariant spectator quark model 
and CBM are equivalent in the SU(2) symmetric limit,
although in the case of 
the covariant spectator quark model $\mu_q$ 
varies from reaction to reaction.

Thus, we can relate the CBM/SU(6)
results of the bare core (valence quark contributions)
with those of the covariant spectator quark model
defining the quark magnetic moments by Eq.~(\ref{eqMUb}).
The quark magnetic moments
of the covariant spectator quark model
generalize the {\it usual} magnetic moments
by the inclusion of the
octet and decuplet baryon mass
dependence ($M_B$ and $M_{B'}$).
Therefore, the effective quark
magnetic moment $\mu_q$,
defined by Eq.~(\ref{eqMUb}),
differs from transition to transition
in the covariant spectator quark model.
However, as mentioned already,
the {\it familiar} expression is recovered
in the limit $M_{B'}= M_{B}= M_N$,
apart from some constants.

Another interesting point is the dependence
of $\mu_q$ on the overlap integral ${\cal I}(0)$,
which is a consequence of the difference
between the octet and decuplet
radial wave functions of the transition.
In a naive picture with $M_B=M_{B'}$,
the octet and decuplet radial wave functions
can be approximated by the same radial wave function
(defined in the same frame) and the overlap
integral would be ${\cal I}(0) =1$.
In the present case as discussed in Ref.~\cite{Oct2Dec},
${\cal I}(0)$ is about 0.8--0.9, depending on the transitions.

A note is in order about the SU(6) result for $G_M^B(0)$,
given by Eq.~(\ref{eqSU6}) for the
$\gamma^\ast N \to \Delta$ reaction.
The numerical result using the experimental value
for $\mu_p$, is $G_M^B(0)= 2.3$
(including the effect of the nucleon and $\Delta$ masses, see footnote 1).
This result, overestimates the relativistic calculations.
Some relativistic calculations
take into account the differences between
the nucleon and $\Delta$ masses 
and also the nonzero momentum of the nucleon
at $Q^2=0$ in the $\Delta$ rest frame.
The Sato-Lee model~\cite{Diaz07a}
for instance
gives $G_M^B(0)=2.05$.
As for the  covariant spectator quark model,
we recall that the model predicts the upper limit of
$G_M^B(0)=2.07$~\cite{NDelta,Oct2Dec},
but in practice this value
is reduced by the overlap of the
nucleon and $\Delta$ radial wave functions, ${\cal I}(0)$,
which is always smaller than unity as already mentioned.
See Appendix B in Ref.~\cite{Oct2Dec} for details.
For the present study it is not important
even if our expressions differ from those of the SU(6)
by a factor.
For example, the factor
$\sqrt{\frac{2}{3}}$ 
may be a consequence of
the relativistic calculation.  
Also the overlap integral ${\cal I}(0)$,
does not appear in the simple SU(6) quark model expressions,
due to the static approximation
[${\cal I}(0) \to 1$].
The important point is to establish the correspondence
between the analytical expressions
in the SU(6) quark model and those of the
spectator formalism consistently.

Next, we comment on the renormalization of the
baryon wave functions.
In the present calculation
we use the decuplet baryon radial wave functions from Ref.~\cite{Omega}
and those of the octet baryons from Ref.~\cite{OctetMedium}.
In these cases the decuplet baryon wave functions
were determined assuming that they have no meson cloud dressing,
while the octet baryon wave functions were determined
assuming a small pion cloud dressing.
As discussed already in Ref.~\cite{Oct2Dec},
the correction due to the renormalization
of the octet baryon wave functions
(due to the pion cloud dressing) is
small, and can be neglected in a first approximation
(less than 4\% effect).
In the present work we include kaon and a eta clouds
in addition to the pion cloud.
Although we cannot calculate
the renormalization effects due to these mesons
for the baryon wave functions
in the covariant spectator quark model framework,
we will assume, as was already done for the pion could, 
that the meson cloud effects are small and can be 
neglected in the normalization
of the wave functions in a first approximation.
Later we will discuss the renormalization
effect due to the meson cloud effects,
since the meson cloud contributions depend also on
the wave functions.

\section{Meson cloud contributions}
\label{secMeson}

To estimate the meson cloud contributions for the processes shown
by the diagrams (a) and (b) in Fig.~\ref{figMesonCloud},
we apply the CBM~\cite{Thomas84}.
As usually practiced in the CBM, we use the static approximation
and neglect the momentum of the baryons in the initial, intermediate
and final baryon states, by replacing the respective energies
by their masses~\cite{Thomas84,Thomas83,TsushimaCBM,Lu97}.
The same approximation is also used in the
heavy baryon chiral perturbation
theory~\cite{Jenkins91,Bernard92,Bernard08,Wang09}.
In addition, we ignore the possible center-of-mass correction
for the 3-quark composite baryon (core) systems,
keeping in mind that this correction reduces
the bare core transition amplitudes for the $\gamma^* N \to \Delta$
reaction by 5 to 10\% in the region $Q^2 \alt 0.5$ GeV$^2$~\cite{Lu97}.
The effects are expected to be even smaller
for the remaining reactions, since the
corresponding baryons are heavier.

Although the approximations discussed above
break the Lorentz covariance,
the phenomenological successes and practices in describing the physics
in the low $Q^2$ region~\cite{Thomas84,Thomas83,TsushimaCBM,Lu97},
suggest that the approximations may be well under control
in the present study, particularly at $Q^2=0$
(small kinematic corrections).

To carry out the calculations of the meson cloud contributions,
all the intermediate states are summed over utilizing
the standard angular momentum algebra in flavor
and spin spaces combined with the Wigner-Eckart
theorem~\cite{Thomas84,Thomas83,TsushimaCBM}.
Thus, the summation is made based on the
SU(6) symmetry at the flavor-spin wave function level. 
The SU(3) breaking effects are, partially included
using the physical baryon and meson masses, and
via the quark masses, $m_u=m_d \ne m_s$.
Note that in the covariant spectator quark model
the  SU(3) symmetry is explicitly broken
in the octet and decuplet baryon wave functions.

The equations derived in the CBM
for $Q^2=0$, depend only on one-dimensional integrals.
In some cases, the CBM integrals have 
singularities in the integrand functions
(poles associated with physical baryons or mesons
in the intermediate states).
These poles yield imaginary parts for the calculated integrals.
For simplicity we evaluate those integrals
using the {\it principal value integral}.
Based on the results from the CBM~\cite{Lu97}
for the $\gamma^* N \to \Delta$ reaction,
we may expect the imaginary part
to be about $15-20$\% of the real part near $Q^2=0$.
Since the decay width depends on $|G_M(0)|^2$,
this approximation has only a small effect in the final results
(a 20\% imaginary part of the real part on $G_M(0)$
leads to a 4\% correction for $|G_M(0)|^2$).

\subsection{Direct coupling with the meson}

We first consider the contributions from the 
processes represented by
the diagram (a) in Fig.~\ref{figMesonCloud}.
In the following the upper index $M$ stands
for the meson ($M=\pi,K$).
The CBM loop integral functions~\cite{Thomas84,Thomas83}
for the initial ($B$) and final ($B'$) baryons 
that depend also on the intermediate baryon $B_1$ states,
for the pion and kaon cloud diagrams,
will be denoted
by $H_{BB'}^\pi(B_1)$ and  $H_{BB'}^K(B_1)$,
respectively.

The contributions from the diagram (a)
for the $\gamma^\ast B \to B'$
can be written as
\ba
G_M^{{\rm MC}a} = \sum_{M,B_1}  C^{M}_{BB'; B_1} H_{BB'}^{M} (B_1),
\label{Hintegral}
\ea
where $C^{M}_{BB'; B_1}$ are the coefficients
calculated in the CBM framework, and are presented
in Appendix~\ref{apDiagA}.

The explicit expression for $H_{BB'}^M(B_1)$ is,
\ba
& &
H_{BB'}^M(B_1)=
\frac{1}{12\pi^2}
\left( \frac{f_{\pi NN}}{m_\pi}\right)^2 \nonumber \\
& &
 \times
\int_0^{\infty} dk
\left\{
\frac{k^4 [j_0(k R) + j_2(k R)]^2}
{\omega_k[4 \omega_k^2-(M_{B'}-M_B)^2]} \right.
\nonumber  \\
& &
\left.
\times
\frac{4 \omega_k + 2 M_{B_1} - M_{B'}-M_B}{
(M_{B_1}-M_B + \omega_k)
(M_{B_1}-M_{B'} + \omega_k)} \right\},
\label{HM1}
\ea
where $R$ is the bag radius,
$j_l$ ($l=0,2$) are the spherical Bessel functions arising
from the CBM form factor, 
$\omega_k= \sqrt{m_M^2 + k^2}$ for $M=\pi,K$ is the meson energy,
and $f_{\pi NN}$ is the pion-nucleon coupling constant.
As already mentioned the integral symbol should be read as
the principal value integral.

In the present work we take a typical,
successful value for the bag radius,
$R=1$ fm~\cite{TsushimaCBM}.
The dependence of the calculated quantities on the values of the
bag radius chosen, can be found in Refs.~\cite{Thomas84,Thomas83,Lu97}.

The factor $\left(\frac{f_{\pi NN}}{m_\pi} \right)^2$
is included in the loop integral definition
for all baryon and meson cases,
since all the couplings are redefined in terms of $f_{\pi NN}$.
For discussions about the renormalized $f_{\pi NN}$ value used in the CBM,
see Ref.~\cite{Thomas83,Lu97}.

\begin{table*}[t]
\begin{tabular}{l   cc cc cc cc cc}
\hline
\hline
    && $G_M^B(0)$ &&
$G_M^{{\rm MCa}\pi}(0)$ &&$G_M^{{\rm MCb}\pi}(0)$ &&  $G_M^{{\rm MC}\pi}(0)$ && $G_M(0)$\\
\hline
\hline
CBM [$\tilde G_M(0)$]  && 1.633 && 0.883 && 0.754  &&
1.634  && 3.270
\\[.1cm]
Spectator [$G_M(0)$]  && 1.633 && 0.713 && 0.610  && 1.323 && 2.956
\\[0.02in]
\hline 
\hline
\end{tabular}
\caption{
Pion cloud contributions for the $\gamma^\ast N \to \Delta$ reaction.
The quantities with the superscript $\pi$ refer only to the pion cloud. 
The first entry
includes the results from the CBM,
while the second 
includes the corresponding quantities from the
covariant spectator quark model.} 
\label{tableNDelta}
\end{table*}

\subsection{Coupling with intermediate baryon states}

Next, we consider the contributions from the diagram (b),
due to the clouds of the pion, kaon and $\eta$ meson.
Since the processes depend
on the intermediate baryon states $B_1$ and $B_2$,
the respective contributions generally depend
on the intermediate state transition form factors
between $B_1$ and $B_2$,
and these can, in the SU(6) quark model, be
represented by the combinations of the quark
magnetic moments $\mu_q$.

In order to obtain a simple estimate
for the meson cloud contributions without
explicitly summing over a huge number of the intermediate states,
we use a technique developed and used in
the CBM framework~\cite{TsushimaCBM}
with the exact isospin symmetry, $\mu_u = \mu_d$.
The use of the isospin symmetry simplifies the calculation
drastically by reducing the number of terms to be considered,
and can be justified when
the difference between $\mu_u$ and $\mu_d$ is small.

In the following calculations of the meson cloud effects
we will replace the CBM quark magnetic moments by
these of the covariant spectator quark model
as defined by Eq.~(\ref{eqMUb}).
In order to keep the isospin symmetry in those
calculations we replace $\mu_u$ and $\mu_d$
by an average $\bar \mu_u$ to be defined later.

Then, the contributions from the  diagram (b)
for the  $\gamma^\ast B \to B'$ transition
can be written as
\ba
G_M^{{\rm MBb}} = \sum_{M,B_1,B_2} D_{BB';B_1 B_2}^M H_{BB'}^{2M}(B_1,B_2),
\ea
where the CBM-based integral is represented by $H^{2M}_{BB'}$
to be defined next,
and $D_{BB';B_1 B_2}^M $ are the coefficients
which depend on the effective magnetic moments $\mu_q$.
The expressions for $D_{BB';B_1 B_2}^M $
are given in Appendix~\ref{apDiagA}.

The integral $H^{2M}_{BB'}$ is defined by
\ba
& &
H_{BB'}^{2M}(B_1,B_2)=
\frac{1}{12\pi^2}
\left( \frac{f_{\pi NN}}{m_\pi}\right)^2 \nonumber \\
& &
 \times
\int_0^{\infty} dk
\left\{
\frac{k^4 [j_0(k R) + j_2(k R)]^2}
{\omega_k
(M_{B_1}-M_B + \omega_k)
(M_{B_2}-M_{B'} + \omega_k)} \right\}.
\label{H2} \nonumber \\
& &
\ea
Again, the principal value integration should be understood.
In Eq.~(\ref{H2})
$B_1$ and $B_2$ are the baryons in the intermediate
states with masses $M_{B_1}$ and $M_{B_2}$, respectively,
and the upper index $2M$ indicates that
there are two baryon propagators while a meson $M$
is in the air.
Besides that, the functions are obtained with a static
approximation, the same as for the $H_{BB'}^M(B_1)$ case,
but we also have now 
$\omega_k= \sqrt{m_\eta^2 + k^2}$ when $M=\eta$.

To be consistent with the SU(2) symmetry
in the calculation of the function $D_{BB';B_1 B_2}^M$
for the meson cloud contributions,
we replace in the expressions for the
magnetic moments, $\mu_u$ and $=\mu_d$,
by the average,
\ba
\bar \mu_u \equiv \frac{1}{3}(2\mu_u + \mu_d).
\ea
With this definition,
the results for the core contributions
are the same for the CBM and the
covariant spectator quark model
for the reactions
$\gamma^\ast N \to \Delta$
and $\gamma^\ast \Lambda \to \Sigma^{* 0}$.
The same expression will be used
for the calculation of the meson cloud
contributions represented by the diagram (b).

As for the reactions involving the $\Sigma$ and $\Xi$
in the calculation of the meson cloud effects,
we use the replacement suggested by Eq.~(\ref{eqFV}),
\ba
& &
\frac{1}{6} (2 \mu_u  -\mu_d + 2 \mu_s) 
+
\frac{1}{6} (2 \mu_u + \mu_d)  t_3 \nonumber \\
& &
\to
\frac{1}{6} (\bar \mu_u  + 2 \mu_s) 
+
\frac{1}{2} \bar \mu_u  t_3,
\label{muSX}
\ea
where in the last line
we have replaced
$\mu_u$ and $\mu_d$ by $\bar \mu_u$.
In practice the difference between $\mu_u,\mu_d$
and $\bar \mu_u$ is smaller than 10\%.

The effect of the baryon wave function
renormalization due to the
meson cloud represented by the diagrams (a) and (b),
can be absorbed in the renormalized
coupling constant $f_{\pi NN}$ used in
the CBM~\cite{Lu97}.

\begin{table*}[t]
\begin{tabular}{l   cc cc cc cc cc}
\hline
\hline
    && $G_M^B(0)$ &&  $G_M^{{\rm MCa}}(0)$
  &&$G_M^{{\rm MCb}}(0)$ &&  $G_M^{{\rm MC}}(0)$ && $G_M(0)$\\
\hline
\hline
$\gamma^\ast N \to \Delta$
      && 1.633 && 0.713 && 0.610  && 1.323 &&  2.956
\\[0.02in]
  &&  && 0.017 && 0.037  && &&
\\[0.02in]
  &&  &&       && 0.0062  && &&
\\[0.02in]
  &&  &&  {\bf 0.730}     && {\bf 0.652}  && {\bf 1.383} && {\bf 3.016}
\\[.3cm]
$\gamma^\ast \Lambda \to \Sigma^{* 0}$ &&
1.683 && 0.669  && 0.358 && 1.027 && 2.710 \\[0.02in]
 &&   && 0.068  && 0.289    &&     &&     \\[0.02in]
 &&   &&        && 0.016    &&         \\[0.02in]
 &&   && {\bf 0.737}  && {\bf 0.663} && {\bf 1.400}   &&
{\bf 3.083}    \\[.3cm]
$\gamma^\ast \Sigma^+ \to \Sigma^{* +}$ &&
     2.094 && 0.149  && 0.513 && 0.663 && 2.757 \\[0.02in]
 &&        && 0.155   && 0.269    &&     &&     \\[0.02in]
 &&        &&         && 0.043    &&         \\[0.02in]
 &&   && {\bf 0.304}  && {\bf 0.825} && {\bf 1.129}   &&
{\bf 3.224}    \\[.3cm]
$\gamma^\ast \Sigma^0 \to \Sigma^{* 0}$ &&
     0.969 && 0.000   && 0.270    &&   0.270 && 1.239 \\[0.02in]
 &&        && 0.104   && 0.010    &&     &&     \\[0.02in]
 &&        &&         && 0.015    &&         \\[0.02in]
 &&   && {\bf 0.104}  && {\bf 0.387} && {\bf 0.490}   &&
{\bf 1.460}    \\[.3cm]
$\gamma^\ast \Sigma^- \to \Sigma^{* -}$ &&
    $-0.156$\; \sp && $-0.149$\;\sp   && 0.026    
&&   $-0.124$\; \sp && $-0.279$ \sp  \\[0.02in]
 &&        && 0.052   && $-0.065$\;\sp    &&     &&     \\[0.02in]
 &&        &&         && $-0.012$\; \sp    &&         \\[0.02in]
 &&   &&  $\mathbf{-0.097}$\;\sp  && $\mathbf{-0.052}$\; \sp
 && $\mathbf{-0.149}$  \;\sp &&
$\mathbf{-0.305}$  \;\sp  \\[.3cm]
$\gamma^\ast \Xi^0 \to \Xi^{* 0}$ &&
     2.191 && 0.222   && 0.086    &&   0.308 && 2.499 \\[0.02in]
 &&        && 0.187   && 0.519    &&     &&     \\[0.02in]
 &&        &&         && 0.086    &&         \\[0.02in]
 &&   && {\bf 0.410}  && {\bf 0.691} && {\bf 1.101}   &&
{\bf 3.291}    \\[.3cm]
$\gamma^\ast \Xi^- \to \Xi^{* -}$ &&
    $-0.168$\;\sp && $-0.222$\;\sp  && 0.084    &&   $-0.138$\;\sp && 
$-0.306$\; \sp \\[0.02in]
 &&        &&   0.038   && $-0.108$ \;\sp   &&     &&     \\[0.02in]
 &&        &&           && $-0.0034$  \sp    &&         \\[0.02in]
 &&   &&  $\mathbf{-0.185}$ \;\sp && $\mathbf{-0.028}$ \! \sp
&& $\mathbf{-0.213}$ \!\sp  &&
$\mathbf{-0.380}$  \! \sp  \\[0.02in]
\hline 
\hline
\end{tabular}
\caption{
Meson cloud contributions for the octet to decuplet transition magnetic moments.
In each group the first line indicates the pion cloud contributions,
the second line the kaon cloud,
the third line the eta cloud, and the fourth
the sum of all meson cloud contributions (boldface).
The column $G_M^B(0)$
presents the contributions from the valence quark core.
The column $G_M^{\rm MC}(0)$ and $G_M(0)$
show respectively the total meson cloud contributions
and the final results (both boldface).
The results in the first line for $G_M^{\rm MC}(0)$
and $G_M(0)$ include only the pion cloud contributions.
}
\label{tableMC}
\end{table*}

\section{Results}
\label{secResults}

Before presenting the results, we recall
that the contributions from the valence quark core
are given by the covariant spectator quark
model as discussed in Sec.~\ref{secValence}.
The formalism was discussed in detail in Ref.~\cite{Oct2Dec}.
The important point to recall
is that for $Q^2=0$ the valence quark contributions $G_M^B$
depend only on the quark anomalous moments
$\kappa_u,\kappa_d,\kappa_s$ and
the octet/decuplet radial wave functions
through ${\cal I}(0)$.
The corresponding parameters
were fixed in the previous works.

We start to present the results
by discussing the pion cloud contributions
for the $\gamma^\ast N \to \Delta$ reaction,
and explain how the meson cloud contributions
are calibrated by this reaction.
Next, we will present the results
for all meson cloud contributions
for the $\gamma^\ast B \to B'$ reactions,
and discuss the final results
of the form factor $G_M$, and decay widths $\Gamma$.
Finally, we will compare our results with those
existing in the literature.

\begin{table*}[t]
\begin{center}
\begin{tabular}{l  cc  cc   cc  cc cc}
\hline
\hline
     &&  \sp\sp$G_M(0)$   &&  $|G_M(0)|_{\rm exp}$  &&
$\Gamma ({\rm keV}) $ && $\Gamma_{\rm exp} ({\rm keV})$
\\ 
\hline
$\Delta \to \gamma N$ &&  3.02 && \sp\sp$3.04\pm0.11$ \cite{PDG}
&& 648 && $660\pm47$ \cite{PDG}  \\[.15cm]
$\Sigma^{\ast 0} \to \gamma \Lambda$ && 3.08   &&
\sp\sp$3.35\pm0.57$ \cite{PDG} && 399 && $470\pm 160$ \cite{PDG} \\
  &&    &&
\sp\sp$3.26\pm0.37$ \cite{Keller11a,Keller11b} &&  && $445\pm 102$
\cite{Keller11a,Keller11b} \\[.15cm]
$\Sigma^{\ast +} \to \gamma \Sigma^+ $ && 3.22 &&
\sp\sp$4.10\pm0.57$
\cite{Keller11a}  && 154 && $250\pm70$ \cite{Keller11a} \\
$\Sigma^{\ast 0} \to \gamma \Sigma^0 $ && 1.46 &&  && 32 && \\
$\Sigma^{\ast -} \to \gamma \Sigma^-$ && $-0.31$ \sp
&&   $< 0.8$ \cite{Molchanov04} && 1.4 && $< 9.5$  \cite{Molchanov04} 
\\ [.15cm]
$\Xi^{\ast 0} \to  \gamma \Xi^0 $ && 3.29  &&  && 182 && \\
$\Xi^{\ast -} \to  \gamma \Xi^- $ && $-0.38$ \sp && && 2.4 &&\\
\hline
\hline
\end{tabular}
\end{center}
\caption{Results for $G_M(0)$ corresponding to
the $B' \to \gamma B$ decays. The values for
$|G_M(0)|_{\rm exp}$ are estimated by Eq.~(\ref{eqGamma})
using the experimental values of $\Gamma_{B' \to \gamma B}$.}
\label{tableGM0}
\end{table*}

\subsection{Pion cloud contributions for the
$\gamma^\ast N \to \Delta$ reaction}

The results for the pion cloud contribution arising from the diagrams
(a) and (b), from the CBM,
are presented
in Table~\ref{tableNDelta}  (entry CBM).
As we can see in Table~\ref{tableNDelta},
the final contribution from the pion cloud in this case is 1.634,
which is larger than the estimate made by the covariant
spectator quark model of 1.323 by about 33\%
(see entry Spectator).
This is not surprising, since the CBM tends 
to overestimate the effect of the pion cloud 
for the $\gamma^\ast N \to \Delta$ reaction.
Indeed, in Ref.~\cite{Lu97} the pion cloud gives a
contribution of about 66\% of the total,
a contribution substantially larger than in
the other calculations~\cite{Burkert04a,NDelta,Diaz07a}.

On the other hand, the pion cloud contribution
in the covariant spectator quark model were determined
by a fit to the $\gamma^\ast N \to \Delta$ data
for $Q^2 \le 6 $ GeV$^2$,
combined with the estimate made for the quark core
contribution extracted by the 
Excited Baryon  Analysis Center model~\cite{Diaz07a}
by removing the meson cloud contributions.
In addition, the estimate of the quark core contributions
from the covariant spectator quark model
was compared successfully with the
results of lattice QCD simulations using the
pion masses around 350--650 MeV, where
the pion cloud contribution is expected to be
suppressed.
To compare with the lattice QCD data, the model was generalized
to the lattice QCD regime using
the vector meson dominance parametrization
for the quark current.
See details in Refs.~\cite{Lattice,LatticeD,Omega,OctetFF}.
All these results show that the covariant spectator quark model
provides a robust description of both
the physical and lattice QCD data,
and that it is probably more appropriated
than the CBM parametrization
for the present study.
The small deviation from the result for $Q^2=0$
(bare plus pion cloud) given by the
covariant quark model, 2.96,
compared to the experimental result of $3.02\pm0.03$~\cite{Tiator01},
is a consequence of the global fit
of the covariant spectator quark model for $Q^2 \le 6$ GeV$^2$,
instead of fitting only to the low $Q^2$
region data.

In the following we will use
$\tilde G_M$ to represent the CBM result, and $G_M$
for the present model (covariant spectator quark model).
Also to distinguish between the different
$\gamma^\ast B \to B'$ reactions,
we will use the argument $BB'$ as $G_M(BB')$.
Recall that we are only discussing
the form factors at $Q^2=0$.

\subsection{Meson cloud contributions for the
$\gamma^\ast B \to B'$ reactions}

In order to keep the successful features of the pion cloud
contributions estimated in the covariant spectator quark model
for the $\gamma^\ast N \to \Delta$ transition,
we normalize the CBM result
for the pion cloud $\tilde G_M^{{\rm MC}\pi}(0)$
by the result of the covariant spectator quark model
$G_M^{{\rm MC} \pi}(0) = 3\lambda_\pi = 1.32$,
given by,
\ba
G_{M}^{{\rm MC} \pi} (N\Delta) = {\cal R} \, \tilde G_M^{{\rm MC} \pi} (N\Delta),
\ea
where
\ba
{\cal R}= \frac{3 \lambda_\pi}{ \tilde G_M^{ {\rm MC} \pi} (N\Delta)}.
\ea
Numerically it gives ${\cal R} \simeq 0.81$.

To estimate the effect of the other meson clouds,
the kaon and $\eta$ meson
in the $\gamma^\ast N \to \Delta$ reaction,
and also for the other octet to decuplet transitions,
we use a similar relation, since all the couplings
are related with the coupling constant $f_{\pi NN}$.
Thus, we use in general,
\ba
G_{M}^{{\rm MC}} (BB') = {\cal R} \, \tilde G_M^{{\rm MC}} (BB').
\ea
Except for the fact that we now include the
diagram (b), the procedure is the same as the one used
in the previous work~\cite{Oct2Dec}.
From Eq.~(\ref{eqFBBp}), we get
\ba
f_{BB'}=
\frac{\tilde G_M^{{\rm MC}}(BB')}{ \tilde G_M^{{\rm MC}\pi}(N\Delta)}.
\ea

The results of the meson cloud contributions
from the diagrams (a) and (b) for the octet
to decuplet electromagnetic transition
form factors at $Q^2=0$, are presented in
Table~\ref{tableMC}.
Since we expect the results
for $\gamma^\ast n \to \Delta^0$
and $\gamma^\ast p \to \Delta^+$
to be the same in the present approach,
we use the label $\gamma^\ast N\to \Delta$
to represent both reactions.

In Table~\ref{tableMC} we can see the
contributions from each meson, $\pi,K$ or $\eta$.
We can conclude that the pion cloud
indeed gives the dominant meson cloud contribution
for the $\gamma^\ast N \to \Delta$ reaction.
However, for the other reactions, particularly the kaon cloud,
can give important contributions.
The magnitude of the kaon plus eta cloud
contributions can be obtained by subtracting
the result of the first line (only pion cloud effects)
from the last line (bold, total) for $G_M(0)$.
We can then conclude that the kaon and eta cloud corrections
are about 0.4 for $\gamma^\ast \Lambda \to \Sigma^{* 0}$
and 0.5 for $\gamma^\ast \Sigma^+ \to \Sigma^{* +}$.
%

\begin{table*}[t]
\begin{tabular}{l   cc cc cc}
\hline
\hline
      && $\Delta \to \gamma N$ &&
  $\Sigma^{*0} \to \gamma \Lambda$ &&
$\Sigma^{*+} \to \gamma \Sigma^+$ \\
\hline
U-spin \cite{Lipkin73,Oct2Dec} && && $292\pm27$ && $138\pm13$ \\
HB$\chi$PT \cite{Butler93a}
 && 670-790 && 252-540 && 70-220 \\
Alg.~Mod.~\cite{Bijker00} && 342-344 && 221.3 && 140.7 \\
QCD SR~\cite{Wang09} && 887 && 409 && 150 \\
Large $N_c$ \cite{Lebed11} && $669\pm42$ && $336\pm81$ && $149\pm36$ \\
Spectator && 648 && 399 && 154 \\ [.15cm]
Data~\cite{PDG,Keller11a,Keller11b}
&& $660\pm47$ && $470\pm160$ && $250\pm70$ \\
      &&            && $445\pm102$ &&            \\
\hline
\hline
\end{tabular}
\caption{
Results for the $B' \to \gamma B$ decay widths (in keV)
given by several works.}
\label{tableModels}
\end{table*}

The kaon cloud effects in some cases are comparable,
or larger than those of the pion cloud, particularly
for the diagram (b).
See for instance the reactions
$\gamma^\ast \Lambda \to \Sigma^{* 0}$, $\gamma^\ast \Sigma^+ \to \Sigma^{* +}$
and $\gamma^\ast \Xi^0 \to \Xi^{* 0}$.

Globally, the meson cloud contributions
can be about 45\% of the total
for the cases  with $|G_M(0)| \approx 3$,
or even larger for
the $\gamma^\ast \Sigma^{*-} \to \Sigma^-$
and $\gamma^\ast \Xi^{*-} \to \Xi^-$ cases.

Another interesting point is the magnitude
of the contributions from the diagram (b).
They are in most cases similar or larger
than the contributions from the diagram (a).
This is a consequence of two main factors:
i) the quark magnetic moments $\mu_q$
are significant (about 2--3 nuclear magneton), 
which enhances the effect;
ii) the diagram (b) has a
large number of intermediate states to be summed over. 
See Appendix~\ref{apDiagA}.

Note that, this feature contradicts the assumption
made in the previous work~\cite{Oct2Dec},
that the diagram (a) is expected to give
the leading order contribution.
However, since in the previous study
the meson cloud contributions
were normalized by the total pion cloud contribution
of the $\gamma^\ast N \to \Delta$ transition,
the difference between the previous and
the new results 
for all the reactions is not drastic.  
Based on the present results for the
$\gamma^\ast N \to \Delta$ transition,
where roughly 50\% of the meson cloud comes
from diagram (a) and (b), we may
regard the previous result as a consequence
of the assumption that both diagrams have
the same effect (50\%) for all the reactions.
Recall that only the pion was considered in the previous work.

In the present study, we also normalize
the meson cloud contributions
by the pion cloud contribution
for the $\gamma^\ast N \to \Delta$ transition,
but we leave the contributions from the  diagrams
(a) and (b) independent, as can be seen in Table~\ref{tableMC}.

Then, we can conclude that the explicit
inclusion of the contributions form the diagram (b),
increases the contribution of the meson cloud,
and improves the description of the
$\Sigma^{*0} \to \gamma \Lambda$ and
$\Sigma^{*+} \to \gamma \Sigma^+$ data
as will be discussed next.
For further discussion, we recall
that the $\Sigma^{*0} \to \gamma \Lambda$ and
$\Sigma^{*+} \to \gamma \Sigma^+$ decay widths
given in the Particle Data Group (PDG)~\cite{PDG},
were underestimated respectively 1.2 and 2.4 standard deviations
in the previous work~\cite{Oct2Dec}.

The final results for $G_M(0)$
are also presented in Table~\ref{tableGM0},
in comparison with the estimates
extracted from the experimental decay widths~\cite{Oct2Dec}.
The estimates were made assuming the dominance of
$G_M(0)$ compared to the quadrupole electric form factor $G_E(0)$.
From Table~\ref{tableGM0}, one can see that
the present model can describe well the data for
the $\Sigma^{*0} \to \gamma \Lambda$
(less than one standard deviation),
and underestimates the
$\Sigma^{*+} \to \gamma \Sigma^+$ data
1.5 standard deviations.
These features may be regarded as a significant improvement compared to
our previous result and 
other theoretical estimates 
(see discussion in the next section).

Using the model results obtained for $G_M(0)$, we calculate
the decuplet electromagnetic decay widths,
assuming the dominance of $G_M$,
\ba
\Gamma_{B' \to \gamma B}=
\frac{\alpha}{16}
\frac{(M_{B'}^2-M_B^2)^3}{M_{B'}^3 M_B^2} |G_M(0)|^2,
\label{eqGamma}
\ea
where $\alpha=\frac{e^2}{4\pi} \simeq \frac{1}{137}$
is the electromagnetic fine structure constant.
The $G_M$ dominance is a good approximation
according to theoretical estimates
and the experimental results for the
$\gamma^\ast N \to \Delta$.
(See Ref.~\cite{Oct2Dec} for a more detailed discussion.)

Our predictions for the decay width, $\Gamma \equiv \Gamma_{B' \to \gamma B}$,
are also presented in Table~\ref{tableGM0}.
For the cases of $\Delta \to \gamma N$ and
$\Sigma^{* 0} \to \gamma \Lambda$, we also present 
the results from PDG~\cite{PDG}.
In addition we present the results for $\Sigma^{* 0} \to \gamma \Lambda$,
and $\Sigma^{* +} \to \gamma \Sigma^+$
from Refs.~\cite{Keller11a,Keller11b}.
Our model results deviate from the data
only for the $\Sigma^{* +} \to \gamma \Sigma^+$ reaction
by 1.4 standards deviations.
The upper limit for the
$\Sigma^{* -} \to \gamma \Sigma^-$ reaction~\cite{Molchanov04},
is also shown in Table~\ref{tableGM0}.

We call attention to the fact that     
the present estimate of the meson cloud contribution
is affected by some uncertainties related
to the effective quark magnetic moments
used to calculate the diagram (b).
Since some of the intermediate states
correspond to elastic transitions (where ${\cal I}(0)=1$),
we can question the use of the prescription
(\ref{eqMUb}) with the factor  ${\cal I}(0) \le 1$
(octet-decuplet radial wave function overlap integral),
given by the inelastic $\gamma^\ast B \to B'$ reaction.
Therefore, an upper limit for the meson cloud contribution
can be obtained by setting ${\cal I}(0)=1$ (perfect overlap of
the radial wave functions).
In this case the final results for
the decay width are enhanced by 1--6\%.
In particular, the
$\Sigma^{* 0} \to \gamma \Lambda$ decay width increases by 
4.4\% and that for the
$\Sigma^{* +} \to \gamma \Sigma^+$  case in 2.6\%.
Therefore, in the latter case, a possible
enhancement due to the intermediate state
baryon wave function overlaps
is small, and it does not significantly increase 
the final result to enough to bring the present result
closer to the experimental data.

Taking into account the typical uncertainty in the CBM
of about 10\%, and also assuming that the meson cloud contribution
is about 50\% of the total in the CBM, this gives
about 5\% ambiguity.
Combining the two ambiguities, one from
the wave function overlap of 1--6\%, and
the other from the CBM estimate for meson cloud
contributions of about 5\%,
we can conclude that our estimate can be
affected by a value around 10\%.

Another interesting exercise can be to check if
the discrepancy between our estimate and the
experimental $\Sigma^{* +} \to \gamma \Sigma^+$ decay width,
may be a consequence of neglecting the effect of $G_E(0)$
in the  calculation of the decay width.
In this case, the deviation from the data
would be the result of dropping the term $3 |G_E(0)|^2$
in the sum with $|G_M(0)|^2$ in the decay width calculation.
Using our result for $|G_M(0)|$,
we would be able to reproduce the experimental
$\Sigma^{* +} \to \gamma \Sigma^+$ decay width
if $|G_E(0)|$ is about 30\% of $|G_M(0)|$.

\subsection{Discussion}

In general, most of the existing quark models underestimate the
$\Sigma^{*0} \to \gamma \Lambda$ and
$\Sigma^{*+} \to \gamma \Sigma^+$ decay widths
by about 50\%.
Chiral quark models with mesonic effects
are lso included in this category~\cite{Wagner98,Sharma10,Yu06}.
In those cases they give a $\Delta \to \gamma N$ decay width
of about 400 keV, and smaller than the experimental result
of $660\pm47$ keV.
A more detailed comparison between the model results
and data can be found
in the previous work~\cite{Oct2Dec}.

A better result for the $\Sigma^{*+} \to \gamma \Sigma^+$
decay width is obtained by an algebraic model
of the hadronic structure~\cite{Bijker00}.
However,  the $\Delta$ decay width is again underestimated
(see Table~\ref{tableModels}).

Some calculations give closer values to the experimental results,
e.g., the Heavy Baryon Chiral Perturbation Theory~\cite{Butler93a}
as presented in Table~\ref{tableModels}.
The windows associated with the results are however, too broad.

Also, the predictions based on U-spin symmetry
proposed in the seventies~\cite{Lipkin73}, give a good description
of the data, using the updated result for the
$\Delta \to \gamma N$ decay width~\cite{Oct2Dec}.
As can be seen in Table~\ref{tableModels}, the best result
differs, respectively, by 0.8 and 1.4 standard deviations
for the $\Lambda$ and $\Sigma^+$ cases.
It is worth mentioning  that
the estimates made in Ref.~\cite{Keller11a}, also
based on a U-spin symmetry,
have a much better agreement with the data.
However, as also discussed in our previous work~\cite{Oct2Dec},
their U-spin symmetry-based estimates did not take into account
the effect of the baryon masses in the
conversion between the form factors and
the helicity amplitudes.

The results from QCD sum rules~\cite{Wang09}
are close to the
$\Sigma^{*+} \to \gamma \Sigma^+$ and
$\Sigma^{*0} \to \gamma \Lambda$ decay width data,
but overestimate the
$\Delta \to \gamma N$ decay width
by about 220 keV.

An excellent description of all the data
was also obtained in Ref.~\cite{Lebed11}
using a $1/N_c$ expansion.
The unknown coefficients in the expansion
are fitted to the known octet and decuplet
magnetic moments
($n,p,\Sigma^{\pm},\Xi^{0,-}$, $\Delta^+, \Omega^-$
and $\Sigma^0 \to \gamma \Lambda$ transition),
providing a prediction for the remaining cases.
Note that the values from Ref.~\cite{Lebed11}
are very close to our own results.

We would like to emphasize that
our estimate is a pure prediction,
since the parameters involved in the calculations
(quarks anomalous magnetic moments and wave functions)
were already determined and calibrated in the previous works.
The only adjustable ingredient in the present calculation
is the magnitude of the pion cloud
contribution for the $\gamma^\ast N \to \Delta$ reaction,
chosen to match the pion cloud contribution
of the original covariant spectator quark model~\cite{NDeltaD}.

\section{Conclusions}
\label{secConclusions}

In this work we have studied the
decuplet to octet electromagnetic decay widths,
which are 
related to the magnetic transition form factors
defined at $Q^2=0$.
To describe the baryon quark core we
have used the covariant spectator quark model,
the model parameters of which are calibrated in the previous
works on the octet and decuplet baryon systems.
To estimate the effects of the meson cloud,
including those from the pion, kaon and eta meson,
we have been guided by the
cloudy bag model, improved by the result from 
the covariant spectator quark model for the
$\gamma^\ast N \to \Delta$ reaction. 
The effects included as the meson cloud are,
the direct photon coupling to the meson (diagram (a)),
and the photon coupling to the intermediate
baryon states while one meson is in the air (diagram (b)).

We conclude that the inclusion of
the contributions from the diagram (b),
as well as the effects of the kaon cloud
(diagrams (a) and (b)), are both very important.
When the meson cloud contributions are combined with
the quark core contributions
calculated by the covariant spectator quark model,
the present model can reproduce the experimental results well.
The inclusion of only the valence quark contributions,
leads to significant underestimates of the data.
The meson cloud effects are particularly important for the reactions
$\Sigma^{* 0} \to \gamma \Lambda$ and
$\Sigma^{* +} \to \gamma \Sigma^+$.
Furthermore, the effect of the diagram (b) is also
very important for the $\Delta \to \gamma N$ reaction.

In summary, we are able to describe  the
$\Sigma^{* 0} \to \gamma \Lambda$
decay width very well, and also obtain 
a very reasonable result (1.4 standard deviations) for
that of the $\Sigma^{* +} \to \gamma \Sigma^+$.
The present approach also  describes 
the $\Delta \to \gamma N$ decay width rather well.
However, in the last case, the agreement is
a consequence of the fit made previously by the model,
although the explicit inclusion of the extra kaon cloud
effects improves the agreement slightly.

Our predictions for the transition form factors of
the other reactions are consistent with the
estimates made based on the U-spin symmetry, namely,
$G_M (\Sigma^{*+}  \Sigma^+) \approx
G_M (\Xi^{*0}  \Xi^0)$, and
$G_M (\Sigma^{*-}  \Sigma^-) \approx G_M (\Xi^{*-}  \Xi^-)$.

We can, in general, conclude that the meson cloud effects
are of fundamental importance to describe
the $\gamma^\ast B \to B'$ reactions, 
especially in the low $Q^2$ region,
and the decuplet baryon decay widths.
To test further the conclusions of the present study,
accurate experimental determination of the unknown
decuplet baryon electromagnetic decay widths is crucial.
In addition, precise lattice QCD simulations 
for several pion mass values
can also help to constrain the contributions form
the valence quarks, and test our estimates of
the quark core contributions.



\begin{acknowledgments}
The authors would like to thank Y. Kohyama for
the CBM note, which helped the present study.
This work was supported by the Brazilian Ministry of Science,
Technology and Innovation (MCTI-Brazil), and
Conselho Nacional de Desenvolvimento Cient{\'i}fico e Tecnol\'ogico
(CNPq), project 550026/2011-8.
\end{acknowledgments}

\appendix

\section{Meson cloud contributions}
\label{apDiagA}

The meson cloud contributions calculated by the CBM
corresponding to the diagram (a) are presented in  Table~\ref{tableFBBp}.
Compared to the results presented in Ref.~\cite{Oct2Dec}
the present results include a factor $\sqrt{\frac{2}{3}}(2 M_B)$
in each transition.
The factor is necessary to represent 
the form factors in its natural units 
(dimensionless).
Note that, the factor $(2M_B)$ multiplied by
$H_{BB'}^M(B_1)$ gives a dimensionless quantity.

In the exact SU(3) limit, the factors  $\sqrt{\frac{2}{3}}(2 M_B)$
become the same for all the octet to decuplet transitions,
and as a consequence, the factor can be ignored
in the calculation of $f_{BB'}$,
since the factors will be canceled out by the normalization,
divided by the $\gamma^\ast N \to \Delta$ contribution
following the procedure of this work.
Taking the limit $H_{BB'}^\pi (B_1) = H_\pi$
(independent of the octet and decuplet baryon masses)
and  $H_{BB'}^K (B_1) = 0$, we recover the previous results
given in Ref.~\cite{Oct2Dec} for the contributions
from the diagram (a).

\begin{table*}[t]
\begin{center}
\begin{tabular}{l     c }
\hline
\hline
    & $\tilde G_M^{{\rm MCa}}(BB')$ \\  [0.05in]
\hline
$\gamma^* N \to \Delta$  &
$\tilde G_M^{\rm MCa}(N \Delta) = \sfrac{4 \sqrt{2}}{9} (2 M_N)\left[
\frac{1}{5}H_{N\Delta}^\pi(N) +
H_{N\Delta}^\pi(\Delta) +
 \frac{1}{25} H_{N\Delta}^K(\Sigma)
+
\frac{1}{5} H_{N\Delta}^K(\Sigma^\ast)
\right] $  \\ [.2cm]
$\gamma^\ast \Lambda \to \Sigma^*$  &
$\tilde G_M^{\rm MCa} (\Lambda \Sigma^*) = \sfrac{2 \sqrt{2}}{15\sqrt{3}}
(2M_\Lambda)\left[
\frac{4}{5} H_{\Lambda \Sigma^*}^\pi(\Sigma)
+ 4 H_{\Lambda \Sigma^*}^\pi(\Sigma^\ast)
+ \frac{3}{5} H_{\Lambda \Sigma^*}^{K}(N)
-\frac{1}{5} H_{\Lambda \Sigma^*}^{K}(\Xi)
 + 2 H_{\Lambda \Sigma^*}^{K}(\Xi^\ast)
\right] $
 \\ [.3cm]
$\gamma^* \Sigma \to \Sigma^*$
& $\tilde G_M^{\rm MCa}(\Sigma \Sigma^*)
= \sfrac{\sqrt{2}}{3} (2M_\Sigma) \left[
\frac{2}{75} H_{\Sigma \Sigma^*}^K(N) +
\frac{8}{15} H_{\Sigma \Sigma^*}^K (\Delta) +
 \frac{2}{15}
 H_{\Sigma \Sigma^*}^K (\Xi) +
\frac{4}{15}
 H_{\Sigma \Sigma^*}^K (\Xi^\ast)
\right] + $ \\ [.05in]
 &
$\sfrac{\sqrt{2}}{3} (2M_\Sigma)
\left[
\frac{4}{25} H_{\Sigma \Sigma^*}^\pi (\Lambda)
- \frac{8}{75} H_{\Sigma \Sigma^*}^\pi (\Sigma) +
\frac{4}{15} H_{\Sigma \Sigma^*}^\pi (\Sigma^\ast)
 \right. $
\\
 [.05in]
&
$ \left.
- \frac{2}{75} H_{\Sigma \Sigma^*}^K (N) +
 \frac{4}{15}
 H_{\Sigma \Sigma^*}^K (\Delta) +
\frac{2}{15}
 H_{\Sigma \Sigma^*}^K (\Xi) +
\frac{4}{15}
 H_{\Sigma \Sigma^*}^K (\Xi^\ast)
\right] J_3$
\\
 [.3cm]
$\gamma^* \Xi \to \Xi^*$ &
$\tilde G_M^{\rm MCa} ({\Xi \Xi^*})=
\sfrac{\sqrt{2}}{3} (2 M_\Xi)
\left[
- \frac{1}{25} H_{\Xi \Xi^*}^{K}(\Lambda) +
\frac{1}{5} H_{\Xi \Xi^*}^{K}(\Sigma)
+ \frac{2}{5} H_{\Xi \Xi^*}^{K}(\Sigma^\ast)  +
\frac{2}{5} H_{\Xi \Xi^*}^{K}(\Omega)
\right]  + $
 \\  [0.05in]
&
$\sfrac{\sqrt{2}}{3} (2 M_\Xi)
\left[
\frac{4}{75}
H_{\Xi \Xi^*}^\pi(\Xi) +
\frac{4}{15}
H_{\Xi \Xi^*}^\pi(\Xi^*) +
\frac{1}{25}
H_{\Xi \Xi^*}^{K}(\Lambda) +

\frac{1}{15}
H_{\Xi \Xi^*}^{K}(\Sigma)
+
\frac{2}{15}
H_{\Xi \Xi^*}^{K} (\Sigma^\ast)
+
\frac{2}{5} H_{\Xi \Xi^*}^{K}(\Omega)
\right] \tau_3$
\\
 [0.05in]
\hline
\hline
\end{tabular}
\end{center}
\caption{
Meson cloud contributions for $G_M$ from the diagram (a) in Fig.~\ref{figMesonCloud}.
}
\label{tableFBBp}
\end{table*}

To calculate the contributions from the diagram (b) in Fig.~\ref{figMesonCloud},
it is convenient to define the following quantities:
\ba
& &
\mu_S= \sfrac{1}{3}(\bar \mu_u + 2 \mu_s), \nonumber\\
& &
\mu_V= \bar \mu_u, \nonumber\\
& &
\mu_1= \sfrac{1}{3}(2 \bar \mu_u + \mu_s), \nonumber\\
& &
\mu_2 = \bar \mu_u - \mu_s, \nonumber\\
& &
\mu_3= \sfrac{1}{9}(\bar \mu_u + 8\mu_s), \nonumber\\
& &
\mu_4 = \sfrac{1}{3}(-\bar \mu_u + 4 \mu_s). \nonumber\\
\ea
Note that the quantities above are dependent
on the transitions 
under consideration
[see Eq.~(\ref{eqMUb})].

We can now write the meson cloud contributions
corresponding to the diagram (b) as:
\ba
& &\tilde G_M^{\rm MCb}(N\Delta)
=\frac{2 \sqrt{2}}{3} \mu_V  
\nonumber\\
& & 
\times
\left\{
\frac{4}{9}H^{2\pi}_{N\Delta}(N,N) 
+ \frac{5}{9}H^{2\pi}_{N\Delta}(N,\Delta) 
+ \frac{8}{225}H^{2\pi}_{N\Delta}(\Delta,N) 
\right.
\nonumber\\
& &
+\frac{4}{9}H^{2\pi}_{N\Delta}(\Delta,\Delta)
+\frac{4}{25}H^{2K}_{N\Delta}(\Lambda,\Sigma) 
+\frac{1}{5}H^{2K}_{N\Delta}(\Lambda,\Sigma^*)
\nonumber\\
& &
+\frac{8}{225}H^{2K}_{N\Delta}(\Sigma,\Sigma)
+\frac{4}{45}H^{2K}_{N\Delta}(\Sigma^*,\Sigma^*)
-\frac{1}{45}H^{2K}_{N\Delta}(\Sigma,\Sigma^*)
\nonumber\\
& &
\left.
+\frac{4}{225}H^{2K}_{N\Delta}(\Sigma^*,\Sigma)
+\frac{1}{15}H^{2\eta}_{N\Delta}(N,\Delta) \right\},
\ea

\ba
& &\hspace{-2em}\tilde G_M^{\rm MCb}(\Lambda \Sigma^{* 0})
=\sqrt{\frac{2}{3}} \frac{M_\Lambda}{M_N} \mu_V 
\nonumber\\
& &
\hspace{-1em}
\times
\left\{ 
\frac{8}{75}H^{2\pi}_{\Lambda\Sigma^*}(\Sigma,\Lambda)
+\frac{32}{225}H^{2\pi}_{\Lambda\Sigma^*}(\Sigma,\Sigma)
+\frac{8}{45}H^{2\pi}_{\Lambda\Sigma^*}(\Sigma,\Sigma^*)
 \right.
\nonumber\\
& &
+\frac{4}{75}H^{2\pi}_{\Lambda\Sigma^*}(\Sigma^*,\Lambda)
-\frac{8}{225}H^{2\pi}_{\Lambda\Sigma^*}(\Sigma^*,\Sigma)
+\frac{16}{45}H^{2\pi}_{\Lambda\Sigma^*}(\Sigma^*,\Sigma^*)
\nonumber\\
& &
+\frac{4}{15}H^{2K}_{\Lambda\Sigma^*}(N,N)
+\frac{8}{15}H^{2K}_{\Lambda\Sigma^*}(N,\Delta)
+\frac{4}{225}H^{2K}_{\Lambda\Sigma^*}(\Xi,\Xi)
\nonumber\\
& &
+\frac{4}{45}H^{2K}_{\Lambda\Sigma^*}(\Xi,\Xi^*)
+\frac{8}{225}H^{2K}_{\Lambda\Sigma^*}(\Xi^*,\Xi)
+\frac{8}{45}H^{2K}_{\Lambda\Sigma^*}(\Xi^*,\Xi^*)
\nonumber\\
& & \left.
+\frac{8}{75}H^{2\eta}_{\Lambda\Sigma^*}(\Lambda,\Sigma)
\right\},
\ea
\ba
& &\hspace{-2em}\tilde G_M^{{\rm MCb}} (\Sigma \Sigma^\ast)=
\frac{\sqrt{2}}{3}\frac{M_\Sigma}{M_N} \nonumber\\
& &
\hspace{-2em}
\times
\left\{ 
\mu_S\left[
\frac{16}{45}H^{2\pi}_{\Sigma\Sigma^*}(\Sigma,\Sigma^*)
+\frac{8}{225}H^{2\pi}_{\Sigma\Sigma^*}(\Sigma^*,\Sigma) \right.
\right.
\nonumber\\
& &\hspace{-2em}
\left.
+\frac{4}{75}H^{2\eta}_{\Sigma\Sigma^*}(\Sigma^*,\Sigma)
+\frac{4}{9}H^{2K}_{\Sigma\Sigma^*}(\Xi,\Xi^*)
-\frac{8}{225}H^{2K}_{\Sigma\Sigma^*}(\Xi^*,\Xi)
\right]\nonumber\\
& &+\mu_1 \left[
\frac{32}{225}H^{2\pi}_{\Sigma\Sigma^*}(\Sigma,\Sigma)
+\frac{8}{75}H^{2\eta}_{\Sigma\Sigma^*}(\Sigma,\Sigma)
\right]\nonumber\\
& &+\mu_s \frac{8}{75}H^{2\pi}_{\Sigma\Sigma^*}(\Lambda,\Lambda)
-\mu_2 \frac{16}{135}H^{2\pi}_{\Sigma\Sigma^*}(\Sigma^*,\Sigma^*)
\nonumber\\
& &
+\mu_V \left[
  \frac{4}{225}H^{2K}_{\Sigma\Sigma^*}(N,N)
+ \frac{16}{45}H^{2K}_{\Sigma\Sigma^*}(\Delta,\Delta)
\right]\nonumber\\
& &
\left.
+\mu_3 \frac{4}{15}H^{2K}_{\Sigma\Sigma^*}(\Xi,\Xi)
+\mu_4 \frac{8}{45}H^{2K}_{\Sigma\Sigma^*}(\Xi^*,\Xi^*)
\right\} 
\nonumber\\
& &\hspace{-2em}+  
\, J_3 \,
\frac{\sqrt{2}}{3} \frac{M_\Sigma}{M_N} \mu_V \nonumber\\
& &\hspace{-3em} 
\times 
\left\{ 
-\frac{8}{75}H^{2\pi}_{\Sigma\Sigma^*}(\Lambda,\Sigma)
+\frac{4}{15}H^{2\pi}_{\Sigma\Sigma^*}(\Lambda,\Sigma^*)
 \right.
\nonumber\\
& &\hspace{-2em}
+\frac{16}{75}H^{2\pi}_{\Sigma\Sigma^*}(\Sigma,\Lambda)
+\frac{32}{225}H^{2\pi}_{\Sigma\Sigma^*}(\Sigma,\Sigma)
+\frac{8}{45}H^{2\pi}_{\Sigma\Sigma^*}(\Sigma,\Sigma^*)
\nonumber\\
& &\hspace{-2em}
-\frac{4}{75}H^{2\pi}_{\Sigma\Sigma^*}(\Sigma^*,\Lambda)
+\frac{4}{225}H^{2\pi}_{\Sigma,\Sigma^*}(\Sigma^*,\Sigma)
-\frac{8}{45}H^{2\pi}_{\Sigma,\Sigma^*}(\Sigma^*,\Sigma^*)
\nonumber\\
& &\hspace{-2em}
+\frac{4}{45}H^{2K}_{\Sigma\Sigma^*}(N,N)
-\frac{4}{45}H^{2K}_{\Sigma\Sigma^*}(N,\Delta)
+\frac{16}{225}H^{2K}_{\Sigma\Sigma^*}(\Delta,N)
\nonumber\\
& &\hspace{-2em}
+\frac{8}{9}H^{2K}_{\Sigma\Sigma^*}(\Delta,\Delta)
+\frac{4}{45}H^{2K}_{\Sigma\Sigma^*}(\Xi,\Xi)
+\frac{4}{9}H^{2K}_{\Sigma\Sigma^*}(\Xi,\Xi^*)
\nonumber\\
& &\hspace{-2em}-\frac{8}{225}H^{2K}_{\Sigma\Sigma^*}(\Xi^*,\Xi)
-\frac{8}{45}H^{2K}_{\Sigma\Sigma^*}(\Xi^*,\Xi^*)
\nonumber\\
& &\hspace{-2em}
\left.
+\frac{16}{75}H^{2\eta}_{\Sigma\Sigma^*}(\Sigma,\Sigma)
+\frac{4}{75}H^{2\eta}_{\Sigma\Sigma^*}(\Sigma^*,\Sigma)
\right\} ,
\ea
\ba
& &\hspace{-3em}\tilde G_M^{{\rm MCb}} (\Xi\Xi^*)=
\frac{\sqrt{2}}{3} \frac{M_\Xi}{M_N}\nonumber\\
& &\hspace{-3em}
\times \left\{ 
\mu_S \left[
-\frac{1}{15}H^{2\pi}_{\Xi\Xi^*}(\Xi,\Xi^*)
+\frac{4}{75}H^{2\pi}_{\Xi\Xi^*}(\Xi^*,\Xi)
+\frac{1}{5}H^{2\eta}_{\Xi\Xi^*}(\Xi,\Xi^*) \right. \right.
\nonumber\\
& &\hspace{-2em}
\left.
+\frac{4}{75}H^{2\eta}_{\Xi\Xi^*}(\Xi^*,\Xi)
+\frac{2}{3}H^{2K}_{\Xi\Xi^*}(\Sigma,\Sigma^*)
-\frac{4}{75}H^{2K}_{\Xi\Xi^*}(\Sigma^*,\Sigma)
\right]\nonumber\\
& &\hspace{-3em}+\mu_3 \left[
+\frac{2}{25}H^{2\pi}_{\Xi\Xi^*}(\Xi,\Xi)
+\frac{6}{25}H^{2\eta}_{\Xi\Xi^*}(\Xi,\Xi)
\right]\nonumber\\
& &\hspace{-3em}+\mu_4 \left[
\frac{2}{15}H^{2\pi}_{\Xi\Xi^*}(\Xi^*,\Xi^*)
-\frac{2}{15}H^{2\eta}_{\Xi\Xi^*}(\Xi^*,\Xi^*)
\right]
+\mu_1 \frac{4}{15}H^{2K}_{\Xi\Xi^*}(\Sigma,\Sigma)
\nonumber\\
& &\hspace{-3em}
\left.
+\mu_s \left[
\frac{4}{75}H^{2K}_{\Xi\Xi^*}(\Lambda,\Lambda)
+\frac{8}{15}H^{2K}_{\Xi\Xi^*}(\Omega,\Omega)
\right]
+\mu_2 \frac{8}{45}H^{2K}_{\Xi\Xi^*}(\Sigma^*,\Sigma^*)
\right\}\nonumber\\
& &\hspace{-3em}+ \,\tau_3 \,\frac{\sqrt{2}}{3} \frac{M_\Xi}{M_N}\mu_V
\left\{ 
-\frac{2}{225}H^{2\pi}_{\Xi\Xi^*}(\Xi,\Xi)
+\frac{1}{45}H^{2\pi}_{\Xi\Xi^*}(\Xi,\Xi^*) \right.
\nonumber\\
& &\hspace{-2em}
+\frac{2}{45}H^{2\pi}_{\Xi\Xi^*}(\Xi^*,\Xi^*)
-\frac{4}{225}H^{2\pi}_{\Xi\Xi^*}(\Xi^*,\Xi)
-\frac{4}{75}H^{2K}_{\Xi\Xi^*}(\Lambda,\Sigma)
\nonumber\\
& &\hspace{-2em}
+\frac{2}{15}H^{2K}_{\Xi\Xi^*}(\Lambda,\Sigma^*)
+\frac{4}{15}H^{2K}_{\Xi\Xi^*}(\Sigma,\Lambda)
+\frac{16}{45}H^{2K}_{\Xi\Xi^*}(\Sigma,\Sigma)
\nonumber\\
& &\hspace{-2em}
+\frac{4}{9}H^{2K}_{\Xi\Xi^*}(\Sigma,\Sigma^*)
+\frac{4}{75}H^{2K}_{\Xi\Xi^*}(\Sigma^*,\Lambda)
-\frac{8}{225}H^{2K}_{\Xi\Xi^*}(\Sigma^*,\Sigma)
\nonumber\\
& &\hspace{-2em}
+\frac{16}{45}H^{2K}_{\Xi\Xi^*}(\Sigma^*,\Sigma^*)
+\frac{2}{25}H^{2\eta}_{\Xi\Xi^*}(\Xi,\Xi)
+\frac{1}{5}H^{2\eta}_{\Xi\Xi^*}(\Xi,\Xi^*)
\nonumber\\
& &\hspace{-2em}
\left.
+\frac{4}{75}H^{2\eta}_{\Xi\Xi^*}(\Xi^*,\Xi)
+\frac{2}{15}H^{2\eta}_{\Xi\Xi^*}(\Xi^*,\Xi^*)
\right\}   .
\ea

In the equations above the factor $\frac{M_B}{M_N}$
is a consequence of the factor $\sqrt{\frac{2}{3}}(2 M_B)$ combined
with $1/(2M_N)$ in units of the quark magnetic moments.
The final result is thus dimensionless.

\end{document}